\documentclass[a4pfaper,10pt]{article}
\usepackage{amsthm}
\usepackage{times}
\usepackage{graphics}
\usepackage{amssymb,amsbsy,amsmath,amsfonts,amssymb,amscd}
\usepackage{subfigure}

\newtheorem{theorem}{Theorem}[section]

\newtheorem{remark}{Remark\/}

\begin{document}

\title{A mathematical proof of the existence of \\ trends in financial time series}
\author{Michel FLIESS \& C\'{e}dric JOIN \\ ~ \\ ~ \\ INRIA-ALIEN -- LIX (CNRS, UMR 7161)
\\ \'Ecole polytechnique, 91128 Palaiseau, France
\\ {\small{\tt Michel.Fliess@polytechnique.edu}}
\\ ~ \\ \& \\ ~ \\ INRIA-ALIEN --  CRAN (CNRS, UMR 7039) \\
Nancy-Universit\'{e}, BP 239, 54506 Vand\oe{}uvre-l\`es-Nancy, France \\
\small{\tt Cedric.Join@cran.uhp-nancy.fr} }
\date{}

\maketitle

\thispagestyle{empty}


\begin{abstract}
We are settling a longstanding quarrel in quantitative finance by
proving the existence of trends in financial time series thanks to a
theorem due to P. Cartier and Y. Perrin, which is expressed in the
language of nonstandard analysis (\emph{Integration over finite
sets},  F. \& M. Diener (Eds): \emph{Nonstandard Analysis in
Practice}, Springer, 1995, pp. 195--204). Those trends, which might
coexist with some altered random walk paradigm and efficient market
hypothesis, seem nevertheless difficult to reconcile with the
celebrated Black-Scholes model. They are estimated via recent
techniques stemming from control and signal theory. Several quite
convincing computer simulations on the forecast of various financial
quantities are depicted. We conclude by discussing the r\^{o}le of
probability theory.
\end{abstract}

\newpage

\section{Introduction}
Our aim is to settle a severe and longstanding quarrel between
\begin{enumerate}
\item the paradigm of {\em random walks}\footnote{Random walks in finance go
back to the work of Bachelier \cite{bachelier}. They became a
mainstay in the academic world sixty years ago (see, {\em e.g.},
\cite{bernstein1,cootner,merton} and the references therein) and
gave rise to a huge literature (see, {\em e.g.}, \cite{shreve} and
the references therein).} and the related {\em efficient market
hypothesis} \cite{fama} which are the bread and butter of modern
financial mathematics,
\item the existence of {\em trends} which is the key assumption in
{\em technical analysis}.\footnote{Technical analysis (see, {\em
e.g.}, \cite{bechu,kaufman,kirk,muller,murphy} and the references
therein), or {\em charting}, is popular among traders and financial
professionals. The notion of trends here and in the usual time
series literature (see, {\em e.g.}, \cite{gou,hamilton}) do not
coincide.}
\end{enumerate}
There are many publications questioning the existence either of
trends (see, {\em e.g.}, \cite{fama,malkiel,paulos}), of random
walks (see, {\em e.g.}, \cite{lo,taylor}), or of the market
efficiency (see, {\em e.g.},
\cite{stiglitz,rosenberg,taylor}).\footnote{An excellent book by
Lowenstein \cite{lowe} is giving flesh and blood to those hot
debates.}
\\

A theorem due to Cartier and Perrin \cite{cartier}, which is stated
in the language of {\em nonstandard analysis},\footnote{See Sect.
\ref{nsa}.} yields the existence of trends for time series under a
very weak integrability assumption. The time series $f(t)$ may then
be decomposed as a sum
\begin{equation}\label{decomposition}
f(t) = f_{\tiny{\rm trend}}(t) + f_{\tiny{\rm fluctuation}}(t)
\end{equation}
where
\begin{itemize}
\item $f_{\tiny{\rm trend}}(t)$ is the trend,
\item $f_{\tiny{\rm fluctuation}}(t)$ is a ``quickly fluctuating'' function around $0$.
\end{itemize}
The very ``nature'' of those quick fluctuations is left unknown and
nothing prevents us from assuming that $f_{\tiny{\rm
fluctuation}}(t)$ is random and/or fractal. It implies the following
conclusion which seems to be rather unexpected
in the existing literature:\\
\\
{\bf The two above alternatives are not necessarily contradictory
and may coexist for a given time series.}\footnote{One should then
define random
walks and/or market efficiency ``around'' trends.}\\
\\
We nevertheless show that it might be difficult to reconcile with
our setting the celebrated Black-Scholes model \cite{black}, which
is in the heart of the approach to quantitative finance via
stochastic differential equations (see, {\em e.g.}, \cite{shreve}
and the references therein). \\

Consider, as usual in signal, control, and in other engineering
sciences, $f_{\tiny{\rm fluctuation}}(t)$ in Eq.
\eqref{decomposition} as an additive corrupting noise. We attenuate
it, {\em i.e.}, we obtain an estimation of $f_{\tiny{\rm trend}}(t)$
by an appropriate filtering.\footnote{Some technical analysts (see,
{\em e.g.}, \cite{bechu}) are already advocating this standpoint.}
These filters
\begin{itemize}
\item are deduced from our approach to noises via nonstandard analysis
\cite{bruit}, which \begin{itemize} \item is strongly connected to
this work, \item led recently to many successful results in signal
and in control (see the references in \cite{arima}),
\end{itemize}
\item yields excellent numerical differentiation
\cite{mboup}, which is here again of utmost importance (see also
\cite{sm,easy} and the references therein for applications in
control and signal).
\end{itemize}

A mathematical definition of trends and effective means for
estimating them, which were missing until now, bear important
consequences on the study of financial time series, which were
sketched in \cite{coventry}:
\begin{itemize}
\item The forecast of the trend is possible on a ``short'' time interval under
the assumption of a lack of abrupt changes, whereas the forecast of
the ``accurate'' numerical value at a given time instant is
meaningless and should be abandoned.
\item The fluctuations of the numerical values around the trend lead
to new ways for computing standard deviation, skewness, and
kurtosis, which may be forecasted to some extent.
\item  The position of the numerical values above or under the trend may be
forecasted to some extent.
\end{itemize}

The quite convincing computer simulations reported in Sect.
\ref{illus} show that we are
\begin{itemize}
\item offering for technical analysis a sound theoretical basis
(see also \cite{hf,lo2}),
\item on the verge of producing on-line indicators for short
time trading, which are easily implementable on
computers.\footnote{The very same mathematical tools already
provided successful computer programs in control and signal.}
\end{itemize}

\begin{remark}
We utilize as in \cite{coventry} the differences between the actual
prices and the trend for computing quantities like standard
deviation, skewness, kurtosis. This is a major departure from
today's literature where
those quantities are obtained via returns and/or logarithmic
returns,\footnote{See Sect. \ref{tf}.}
and where trends do not play any r\^{o}le.
It might yield a new understanding of ``volatility'', and therefore
a new model-free risk management.\footnote{The existing literature
contains of course other attempts for introducing nonparametric risk
management (see, {\em e.g.}, \cite{ait}).}
\end{remark}

Our paper is organized as follows. Sect. \ref{existence} proves the
existence of trends, which seem to contradict the Black-Scholes
model. Sect. \ref{removal} sketches the trend estimation by
mimicking \cite{easy}. Several computer simulations are depicted in
Sect. \ref{illus}. Sect. \ref{conclusion} concludes by examining
probability theory in finance.

\section{Existence of trends}\label{existence}

\subsection{Nonstandard analysis}\label{nsa}
Nonstandard analysis was discovered in the early 60's by Robinson
\cite{robinson}. It vindicates Leibniz's ideas on ``infinitely
small'' and ``infinitely large'' numbers and is based on deep
concepts and results from mathematical logic. There exists another
presentation due to Nelson \cite{bams}, where the logical background
is less demanding (see, {\em e.g.}, \cite{diener1,diener2,robert}
for excellent introductions). Nelson's approach \cite{nelson-proba}
of probability along those lines had a lasting
influence.\footnote{The following quotation of D. Laugwitz, which is
extracted from \cite{harthong2}, summarizes the power of nonstandard
analysis: {\em Mit \"{u}blicher Mathematik kann man zwar alles gerade so
gut beweisen; mit der nicht-standard Mathematik kann man es aber
verstehen}.} As demonstrated by Harthong \cite{harthong}, Lobry
\cite{lobry0}, and several other authors, nonstandard analysis is
also a marvelous tool for clarifying in a most intuitive way
questions stemming from some applied sides of science. This work is
another step in that direction, like \cite{bruit,arima}.

\subsection{Sketch of the Cartier-Perrin theorem\protect\footnote{The
reference \cite{lobry} contains a well written elementary
presentation. Note also that the Cartier-Perrin theorem is extending
previous considerations in \cite{harthong1,reder}.}}\label{cp}
\subsubsection{Discrete Lebesgue measure and $S$-integrability}
Let $\mathfrak{I}$ be an interval of $\mathbb{R}$, with extremities
$a$ and $b$. A sequence $\mathfrak{T} = \{0 = t_0 < t_1 < \dots <
t_\nu = 1\}$ is called an {\em approximation}  of $\mathfrak{I}$, or
a {\em near interval}, if $t_{i+1} - t_{i}$ is {\em infinitesimal}
for $0 \leq i < \nu$. The {\em Lebesgue measure} on $\mathfrak{T}$
is the function $m$ defined on $\mathfrak{T} \backslash \{b\}$ by
$m(t_{i}) = t_{i+1} - t_{i}$. The measure of any interval $[c, d[
\subset \mathfrak{I}$, $c \leq d$, is its length $d -c$.  The
integral over $[c, d[$ of the function $f: \mathfrak{I} \rightarrow
\mathbb{R}$ is the sum
$$\int_{[c, d[} fdm = \sum_{t \in [c, d[} f(t)m(t)$$
The function $f: \mathfrak{T} \rightarrow \mathbb{R}$ is said to be
$S$-{\em integrable} if, and only if, for any interval $[c, d[$ the
integral $\int_{[c, d[} |f| dm$ is limited and, if $d - c$ is
infinitesimal, also infinitesimal.
\subsubsection{Continuity and Lebesgue integrability}
The function $f$ is said to be $S$-{\em continuous} at $t_\iota \in
\mathfrak{T}$ if, and only if, $f(t_\iota) \simeq f(\tau)$ when
$t_\iota \simeq \tau$.\footnote{$x \simeq y$ means that $x - y$ is
infinitesimal.} The function $f$ is said to be {\em almost
continuous} if, and only if, it is $S$-continuous on $\mathfrak{T}
\setminus R$, where $R$ is a {\em rare} subset.\footnote{The set $R$
is said to be rare \cite{benoit} if, for any standard real number
$\alpha > 0$, there exists an internal set $B \supset A$ such that
$m(B) \leq \alpha$.} We say that $f$ is Lebesgue integrable if, and
only if, it is $S$-integrable and almost continuous.
\subsubsection{Quickly fluctuating functions}
A function $h: \mathfrak{T} \rightarrow \mathbb{R}$ is said to be
{\em quickly fluctuating}, or {\em oscillating}, if, and only if, it
is $S$-integrable and $\int_A h dm$ is infinitesimal for any {\em
quadrable} subset.\footnote{A set is quadrable \cite{cartier} if its
boundary is rare.}

\begin{theorem}\label{theorem}
Let $f: \mathfrak{T} \rightarrow \mathbb{R}$ be an $S$-integrable
function. Then the decomposition \eqref{decomposition} holds where
\begin{itemize}
\item $f_{\tiny{\rm trend}}(t)$ is Lebesgue integrable,
\item $f_{\tiny{\rm fluctuation}}(t)$ is quickly fluctuating.
\end{itemize}
The decomposition \eqref{decomposition} is unique up to an
infinitesimal.
\end{theorem}
$f_{\tiny{\rm trend}}(t)$ and $f_{\tiny{\rm fluctuation}}(t)$ are
respectively called the \emph{trend} and the \emph{quick
fluctuations} of $f$. They are unique up to an infinitesimal.

\subsection{The Black-Scholes model}
The well known Black-Scholes model \cite{black}, which describes the
price evolution of some stock options, is the It\^{o} stochastic
differential equation
\begin{equation}\label{BS}
d S_t = \mu S_t + \sigma S_t dW_t
\end{equation}
where
\begin{itemize}
\item $W_t$ is a standard Wiener process,
\item the {\em volatility} $\sigma$ and the {\em drift}, or {\em trend}, $\mu$ are assumed
to be constant.
\end{itemize}
This model and its numerous generalizations are playing a major r\^{o}le
in financial mathematics since more than thirty years although Eq.
\eqref{BS} is often severely criticized (see, {\em e.g.},
\cite{mandel2,taleb} and the references therein).

The solution  of Eq. \eqref{BS} is the {\em geometric Brownian
motion} which reads
\begin{equation*}\label{bslog}
 S_t = S_0 \exp \left( (\mu - \frac{\sigma^2}{2})t + \sigma W_t \right)
\end{equation*}
where $S_0$ is the initial condition. It seems most natural to
consider the mean $S_0 e^{\mu t}$ of $S_t$ as the trend of $S_t$.
This choice unfortunately does not agree with the following fact:
$F_t = S_t - S_0 e^{\mu t}$ is almost surely not a quickly
fluctuating function around $0$, {\em i.e.}, the probability that $|
\int_{0}^{T} F_\tau d \tau | > \epsilon > 0$, $T
> 0$, is not ``small'', when
\begin{itemize}
\item $\epsilon$ is ``small'', \item $T$ is neither ``small'' nor
``large''.
\end{itemize}
\begin{remark}
A rigorous treatment, which would agree with nonstandard analysis
(see, {\em e.g.}, \cite{al,benoit1}), may be deduced from some
infinitesimal time-sampling of Eq. \eqref{BS}, like the
Cox-Ross-Rubinstein one \cite{cox}.
\end{remark}

\begin{remark}
Many assumptions concerning Eq. \eqref{BS} are relaxed in the
literature (see, {\em e.g.}, \cite{shreve} and the references
therein):
\begin{itemize}
\item $\mu$ and $\sigma$ are no more constant and may be
time-dependent and/or $S_t$-dependent.
\item Eq. \eqref{BS} is no more driven by a Wiener process but by more
complex random processes which might exhibit jumps in order to deal
with ``extreme events''.
\end{itemize}
The conclusion reached before should not be modified, {\em i.e.},
the price is not oscillating around its trend.
\end{remark}

\subsection{Returns}\label{tf}
Assume that the function
$f: \mathfrak{I} \rightarrow \mathbb{R}$ gives the prices of some
financial asset. It implies that the values of $f$ are positive.
What is usually studied in quantitative finance are the {\em return}
\begin{equation}\label{return}
r(t_i) = \frac{f(t_i) - f(t_{i-1})}{f(t_{i-1})}
\end{equation}
and the {\em logarithmic return}, or {\em log-return},
\begin{equation}\label{logreturn}
\mathbf{r}(t_i) = \log (f(t_i)) - \log (f(t_{i-1})) = \log \left(
\frac{f(t_i)}{f(t_{i-1})} \right) = \log \left(1 + r(t_i) \right)
\end{equation}
which are defined for $t_i \in \mathfrak{T} \backslash \{a\}$. There
is a huge literature investigating the statistical properties of the
two above returns, {\em i.e.}, of the time series \eqref{return} and
\eqref{logreturn}.

\begin{remark}
Returns and log-returns are less interesting for us since the trends
of the original time series are difficult to detect on them. Note
moreover that the returns and log-returns which are associated to
the Black-Scholes equation \eqref{BS} via some infinitesimal
time-sampling \cite{al,benoit1} are not $S$-integrable: Theorem
\ref{theorem} does not hold for the corresponding time series
\eqref{return} and \eqref{logreturn}.
\end{remark}

Assume that the trend $f_{\tiny{\mbox{\rm trend}}}: \mathfrak{I}
\rightarrow \mathbb{R}$ is $S$-continuous at $t = t_i$.  Then Eq.
\eqref{decomposition} yields
$$
f(t_i) - f(t_{i-1}) \simeq f_{\tiny{\mbox{\rm fluctuation}}} (t_{i})
- f_{\tiny{\mbox{\rm fluctuation}}} (t_{i-1})
$$
Thus
\begin{equation*}\label{quotient}
r(t_i) \simeq \frac{f_{\tiny{\mbox{\rm fluctuation}}} (t_{i}) -
f_{\tiny{\mbox{\rm fluctuation}}} (t_{i-1})}{f(t_{i-1})}
\end{equation*}

It yields the following crucial conclusion:
\\ ~ \\
{\bf The existence of trends does not preclude, but does not imply
either, the possibility of a fractal and/or random behavior for the
returns \eqref{return} and \eqref{logreturn} where the fast
oscillating function $f_{\tiny{\mbox{\rm fluctuation}}} (t)$ would
be fractal and/or random}.

\section{Trend estimation}\label{removal}
Consider the real-valued polynomial function $x_{N} (t) = \sum_{\nu
= 0}^{N} x^{(\nu)}(0) \frac{t^\nu}{\nu !} \in \mathbb{R}[t]$, $t
\geq 0$, of degree $N$. Rewrite it in the well known notations of
operational calculus (see, {\em e.g.}, \cite{yosida}):
$$
X_N (s) = \sum_{\nu = 0}^{N} \frac{x^{(\nu)}(0)}{s^{\nu + 1}}
$$
Introduce $\frac{d}{ds}$, which is sometimes called the {\em
algebraic derivative} \cite{miku1,miku2}, and which corresponds in
the time domain to the multiplication by $-t$. Multiply both sides
by $\frac{d^\alpha}{ds^\alpha} s^{N + 1}$, $\alpha = 0, 1, \dots,
N$. The quantities $x^{(\nu)}(0)$, $\nu = 0, 1, \dots, N$, which are
given by the triangular system of linear equations, are said to be
{\em linearly identifiable} (see, {\em e.g.}, \cite{arima}):
\begin{equation}\label{triangu}
\frac{d^\alpha s^{N + 1} X_N}{ds^\alpha}  =
\frac{d^\alpha}{ds^\alpha} \left( \sum_{\nu = 0}^{N} x^{(\nu)}(0)
s^{N - \nu} \right)
\end{equation}
The time derivatives, {\em i.e.}, $s^\mu \frac{d^\iota
X_N}{ds^\iota}$, $\mu = 1, \dots, N$, $0 \leq \iota \leq N$, are
removed by multiplying both sides of Eq. (\ref{triangu}) by $s^{-
\bar{N}}$, $\bar{N} > N$, which are expressed in the time domain by
iterated time integrals.

Consider now a real-valued analytic time function defined by the
convergent power series $x(t) = \sum_{\nu = 0}^{\infty} x^{(\nu)}(0)
\frac{t^\nu}{\nu !}$, $0 \leq t < \rho$. Approximating $x(t)$ by its
truncated Taylor expansion $x_{N} (t) = \sum_{\nu = 0}^{N}
x^{(\nu)}(0) \frac{t^\nu}{\nu !}$ yields as above derivatives
estimates.

\begin{remark}
The iterated time integrals are low-pass filters which attenuate the
noises when viewed as in \cite{bruit} as quickly fluctuating
phenomena.\footnote{See \cite{lobry} for an introductory
presentation.} See \cite{mboup} for fundamental computational
developments, which give as a byproduct most efficient estimations.
\end{remark}

\begin{remark}
See \cite{gen} for other studies on filters and estimation in
economics and finance.
\end{remark}

\begin{remark}
See \cite{taylor} for another viewpoint on a model-based trend
estimation.
\end{remark}

\section{Some illustrative computer simulations}\label{illus}
Consider the Arcelor-Mittal daily stock prices from 7 July 1997
until 27 October 2008.\footnote{Those data are borrowed from {\tt
http://finance.yahoo.com/}.}
\subsection{1 day forecast}
\begin{figure*}[!ht]
\centering {{
\rotatebox{-90}{\resizebox{!}{10.5cm}{%
\includegraphics{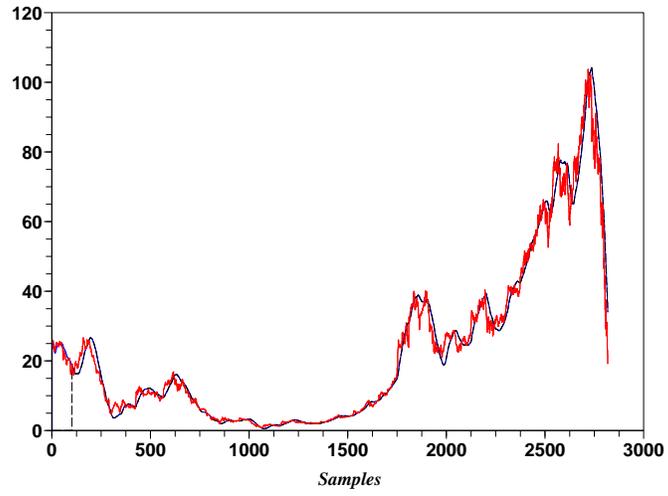}}}}}
\caption{1 day forecast -- Prices (red --), filtered signal (blue
--), forecasted signal (black - -) \label{p1a}}
\end{figure*}

\begin{figure*}[!ht]
\centering {{
\rotatebox{-90}{\resizebox{!}{10.5cm}{%
\includegraphics{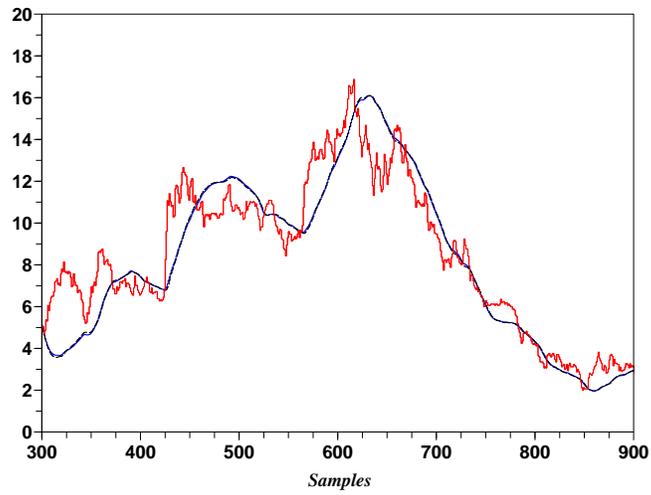}}}}}
\caption{1 day forecast -- Zoom of figure \ref{p1a} \label{p1b}}
\end{figure*}
Figures \ref{p1a} and \ref{p1b} present
\begin{itemize}
\item the estimation of the trend thanks to the methods of Sect. \ref{removal},
with $N = 2$;
\item a $1$ day forecast of the trend by employing a $2^{nd}$-order
Taylor expansion. It necessitates the estimation of the first two
trend derivatives which is also achieved via the methods of Sect.
\ref{removal}.\footnote{Here, as in \cite{coventry}, forecasting is
achieved without specifying a model (see also \cite{sm}).}
\end{itemize}

\begin{figure*}[!ht]
\centering {{
\rotatebox{-90}{\resizebox{!}{10.5cm}{%
\includegraphics{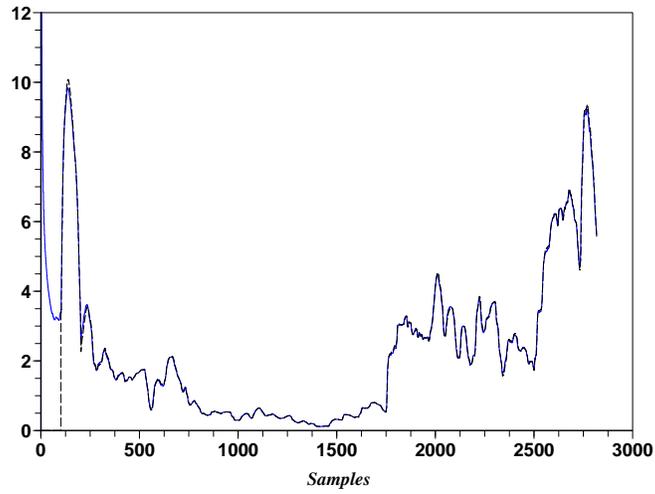}}}}}
\caption{$1$ day forecast -- Standard deviation w.r.t. trend (blue
--), predicted standard deviation (black - -) \label{p1d}}
\end{figure*}

We now look at some properties of the quick fluctuations
$f_{\tiny{\rm fluctuation}}(t)$ around the trend $f_{\tiny{\rm
trend}}(t)$ of the price $f(t)$ (see Eq. \eqref{decomposition}) by
computing moving averages which correspond to various moments
$$
MA_{k,M}(t)=\frac{\sum_{\tau=0}^{M} (f_{\tiny{\rm fluctuation}}(\tau
- M)-\bar{f}_{\tiny{\rm fluctuation}})^k}{M+1}
$$
where
\begin{itemize}
\item $k \geq 2$,
\item $\bar{f}_{\tiny{\rm fluctuation}}$ is the mean of $f_{\tiny{\rm
fluctuation}}$ over the $M + 1$ samples,\footnote{According to Sect.
\ref{cp} $\bar{f}_{\tiny{\rm fluctuation}}$ is ``small''.}
\item $M = 100$ samples.
\end{itemize}
The standard deviation and its $1$ day forecast are displayed in
Figure \ref{p1d}. Its heteroscedasticity is obvious.

\begin{figure*}[!ht]
\centering {{
\rotatebox{-90}{\resizebox{!}{10.5cm}{%
\includegraphics{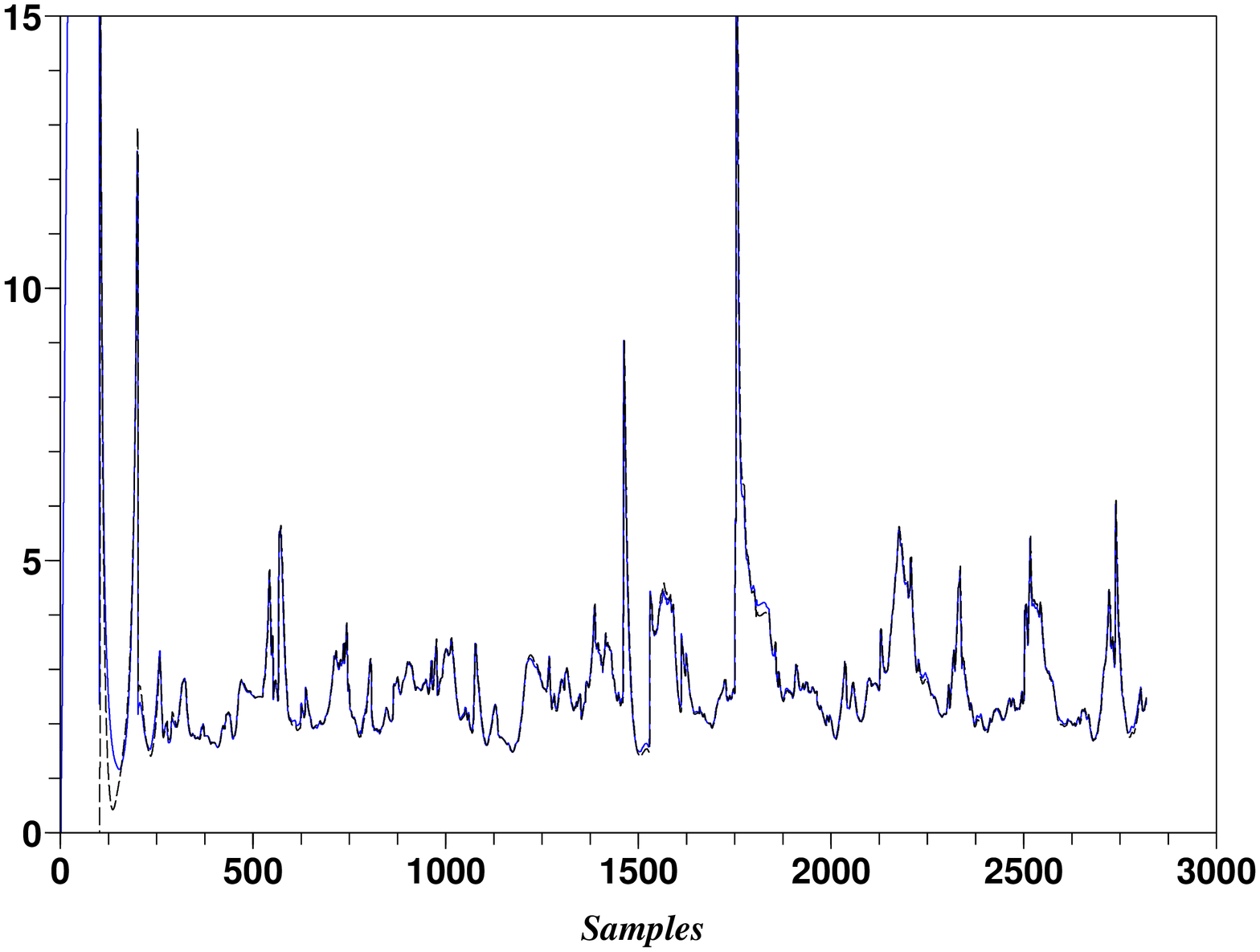}}}}}
\caption{$1$ day forecast -- Kurtosis w.r.t. trend (blue --),
predicted kurtosis (black - -)  \label{p1f}}
\end{figure*}

\begin{figure*}[!ht]
\centering {{
\rotatebox{-90}{\resizebox{!}{10.5cm}{%
\includegraphics{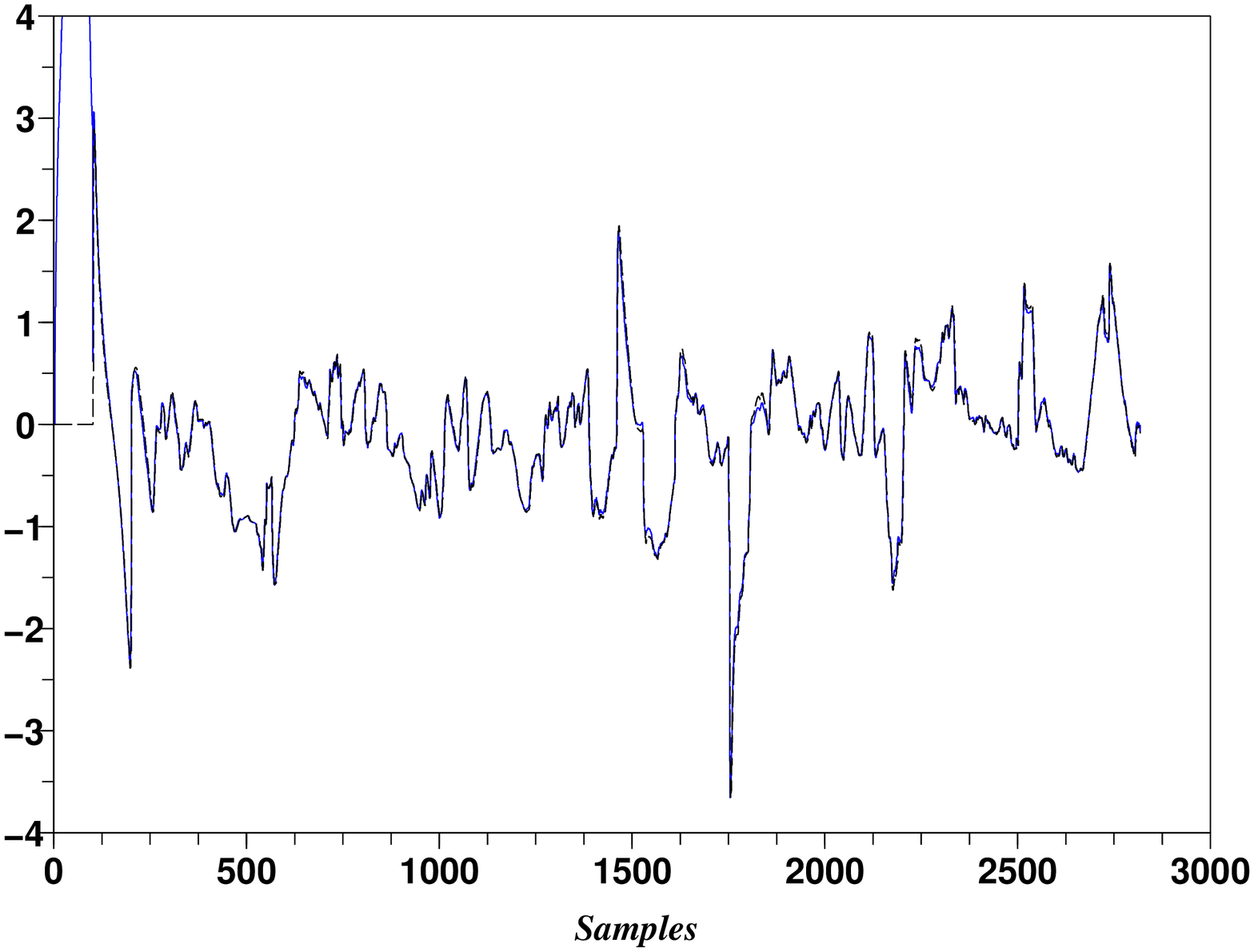}}}}}
\caption{$1$ day forecast -- Skewness w.r.t. trend (blue --),
predicted skewness (black - -)  \label{p1g}}
\end{figure*}

The kurtosis $\frac{MA_{4,100}(t)}{MA_{2,100}(t)^{2}}$, the skewness
$\frac{MA_{3,100}(t)}{MA_{2,100}(t)^{3/2}}$, and their $1$ day
forecasts are respectively depicted in Figures \ref{p1f} and
\ref{p1g}. They show quite clearly that the prices do not exhibit
Gaussian properties\footnote{Lack of spaces prevents us to look at
returns and log-returns.} especially when they are close to some
abrupt change.


\subsection{5 days forecast}

\begin{figure*}[!ht]
\centering {{
\rotatebox{-90}{\resizebox{!}{10.5cm}{%
\includegraphics{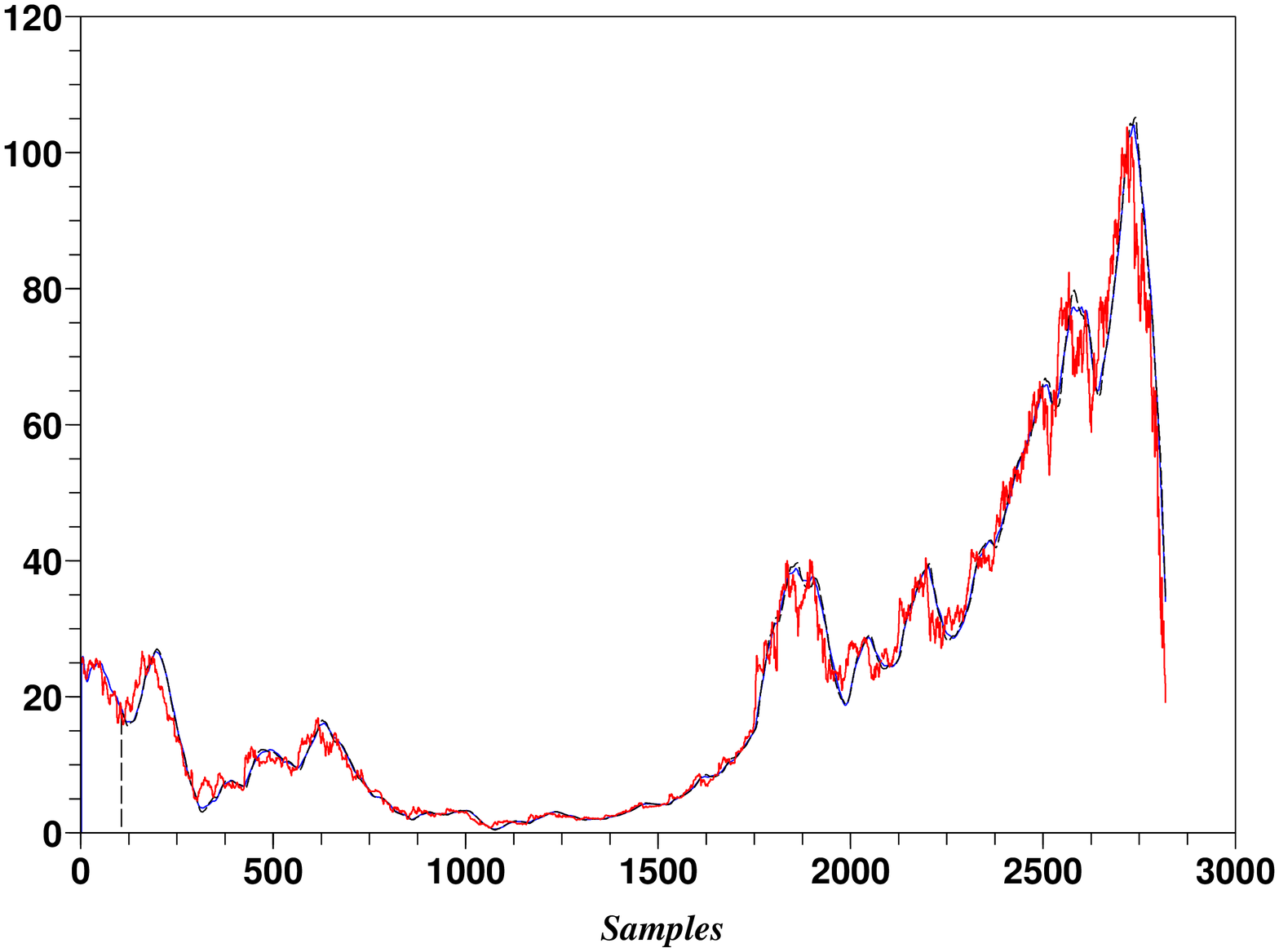}}}}}
\caption{$5$ days forecast -- Prices (red --), filtered signal (blue
--), forecasted signal (black - -)  \label{p2a}}
\end{figure*}
\begin{figure*}[!ht]
\centering {{
\rotatebox{-90}{\resizebox{!}{10.5cm}{%
\includegraphics{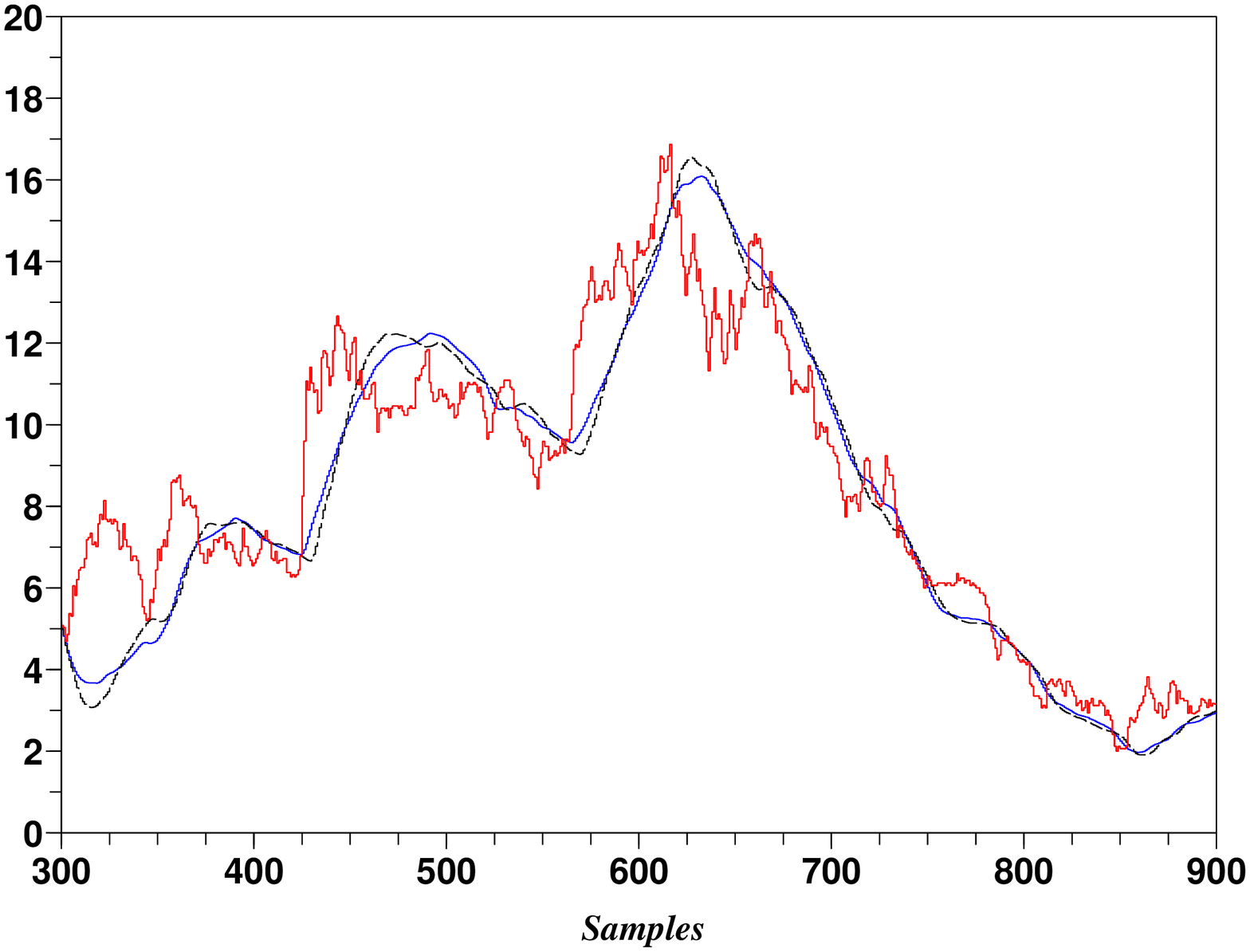}}}}}
\caption{$5$ days forecast --  Zoom of figure \ref{p2a}
\label{p2b}}
\end{figure*}
%
%
\begin{figure*}[!ht]
\centering {{
\rotatebox{-90}{\resizebox{!}{10.5cm}{%
\includegraphics{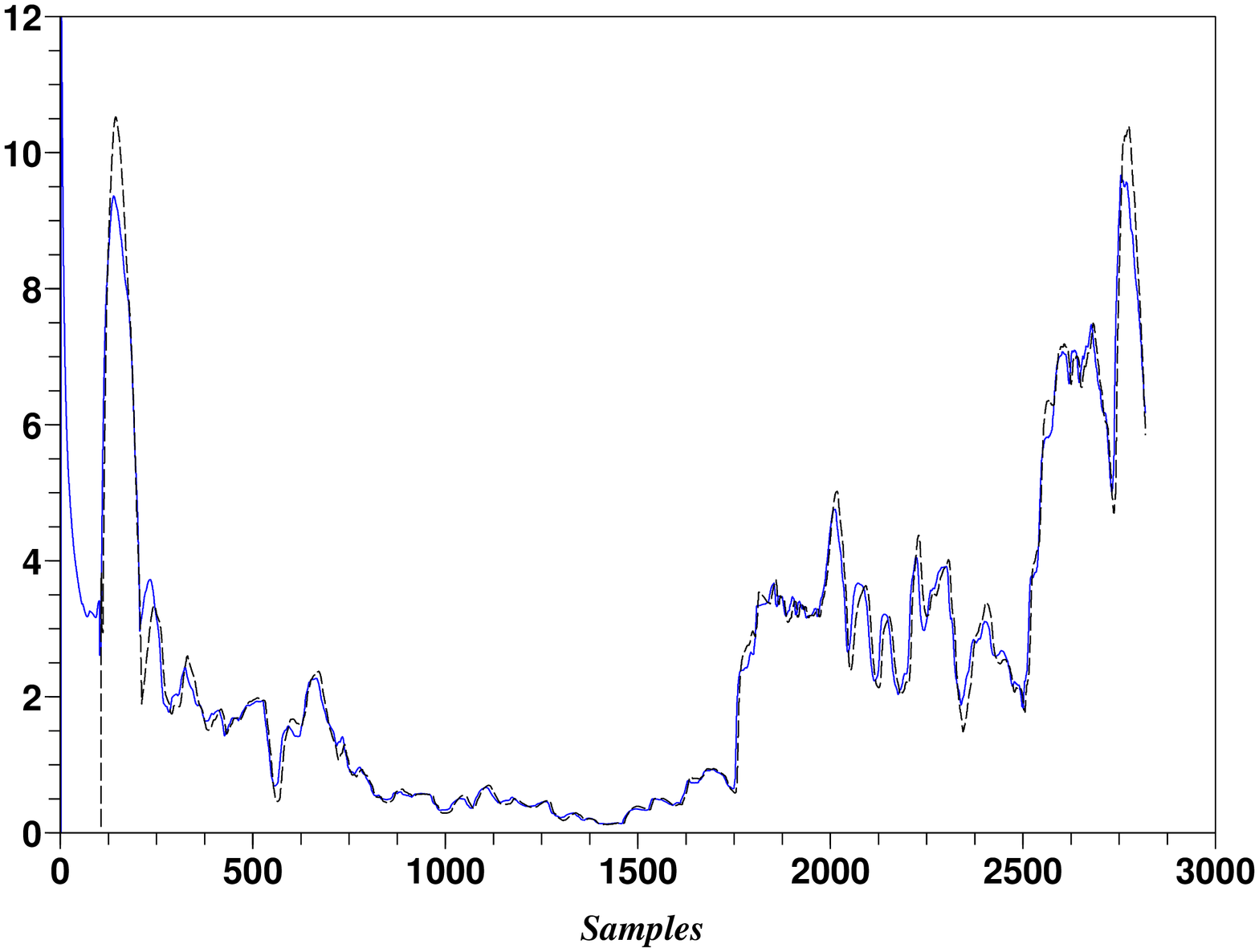}}}}}
\caption{$5$ days forecast -- Standard deviation w.r.t. trend (blue
--), predicted standard deviation (black - -) \label{p2d}}
\end{figure*}
%
%
\begin{figure*}[!ht]
\centering {{
\rotatebox{-90}{\resizebox{!}{10.5cm}{%
\includegraphics{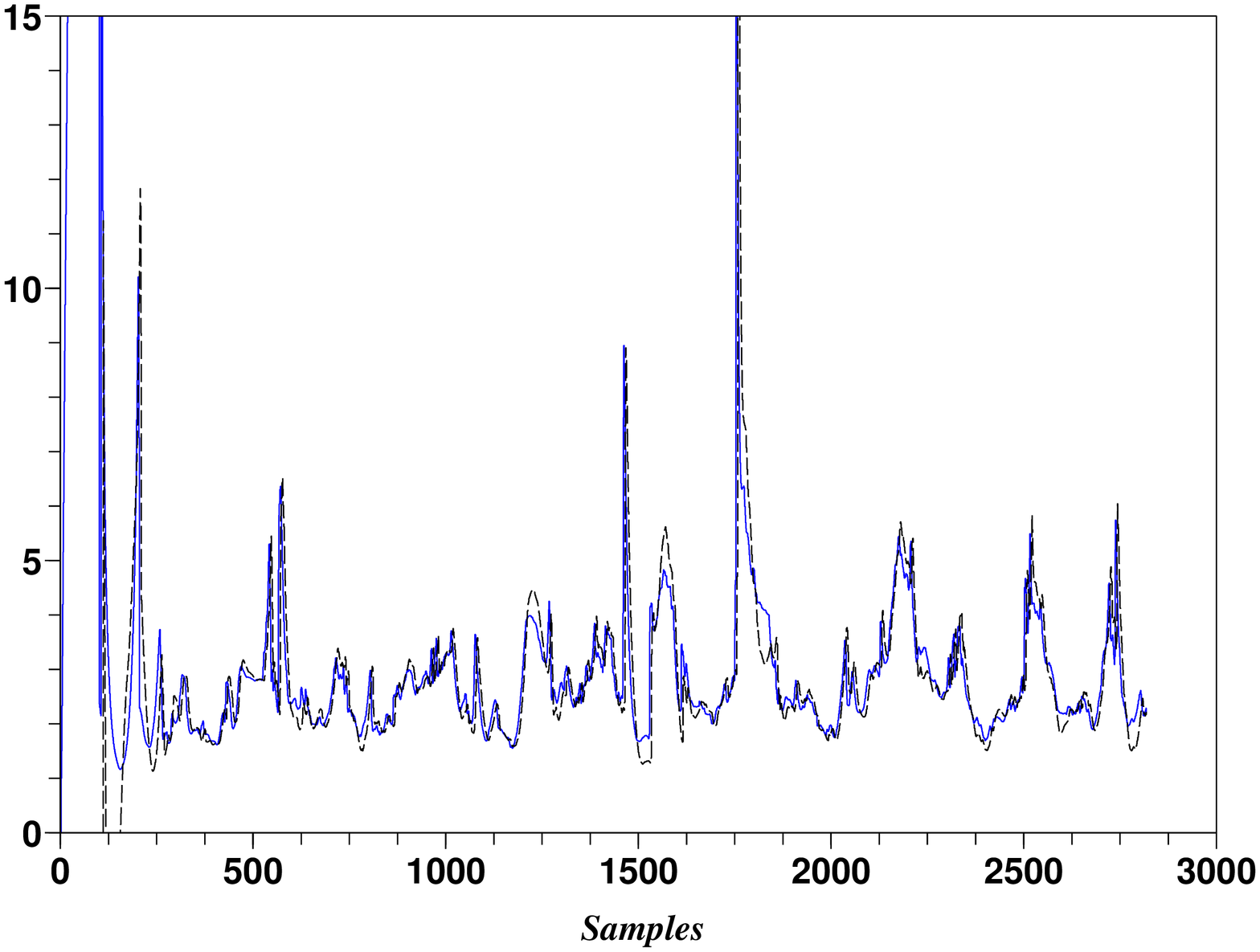}}}}}
\caption{$5$ days forecast -- Kurtosis w.r.t. trend (blue --),
predicted kurtosis (black - -)  \label{p2f}}
\end{figure*}
\begin{figure*}[!ht]
\centering {\subfigure[\footnotesize Skewness of error trend (blue
--), predicted skewness of error trend (black - -) (5 day ahead)]{
\rotatebox{-90}{\resizebox{!}{10.5cm}{%
\includegraphics{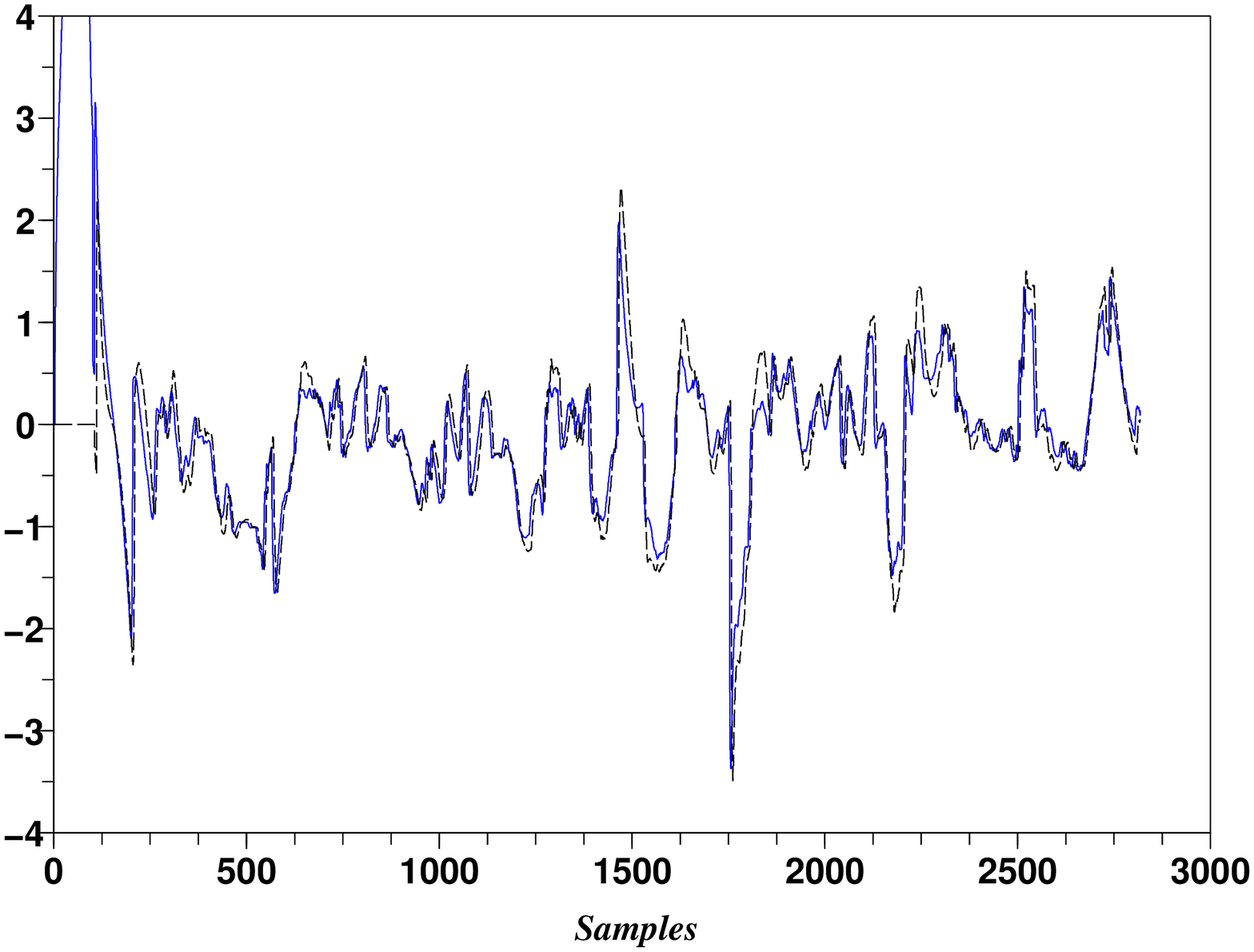}}}}}
\caption{$5$ days forecast -- Skewness w.r.t. trend (blue --),
predicted skewness (black - -) \label{p2g}}
\end{figure*}
%
%

A slight degradation with a $5$ days forecast is visible on the
Figures \ref{p2a} to \ref{p2g}.

\subsection{Above or under the trend?}

\begin{figure*}[!ht]
\centering {{
\rotatebox{-90}{\resizebox{!}{10.5cm}{%
\includegraphics{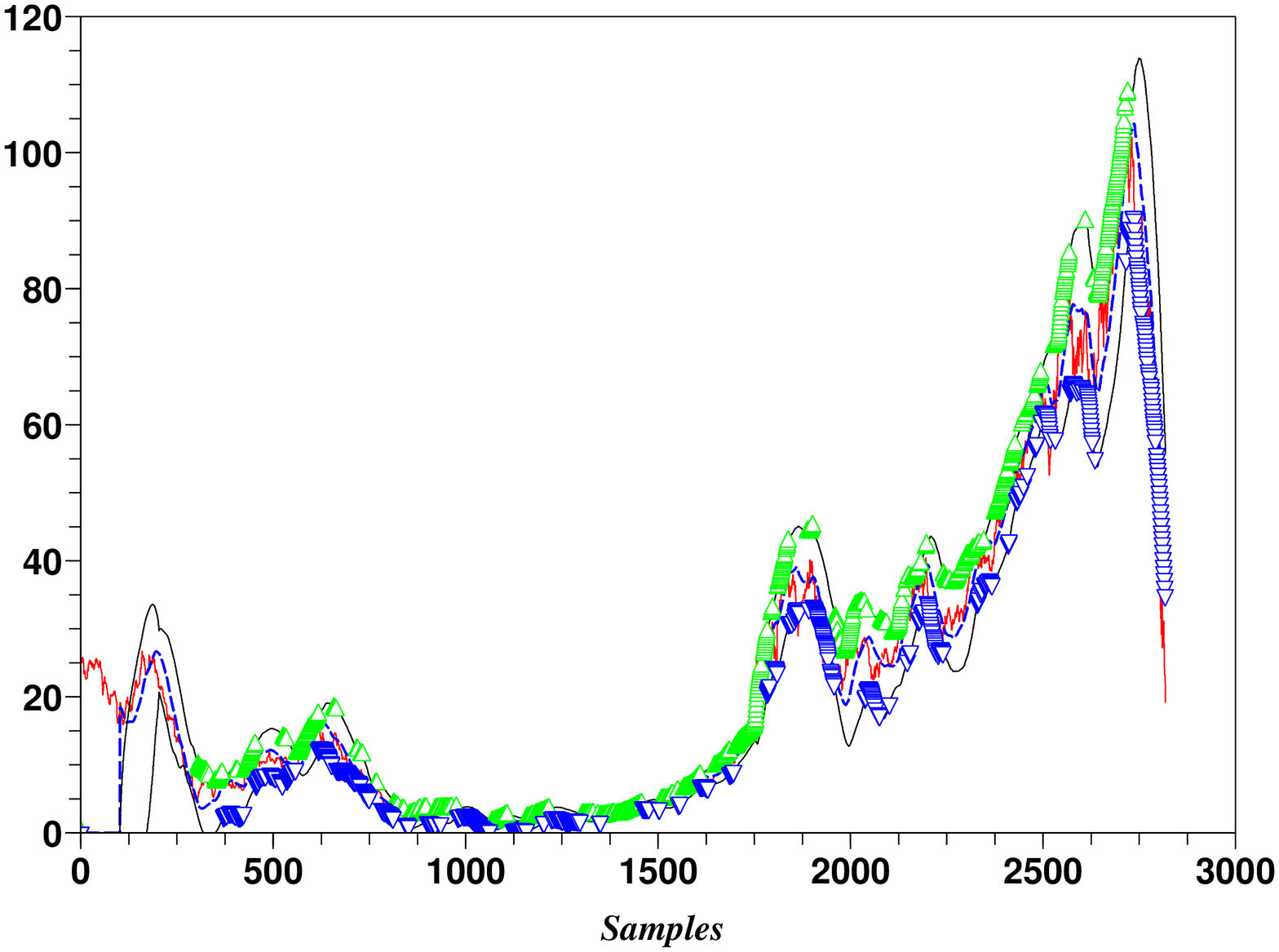}}}}}
\caption{$1$ day forecast -- Prices (red --), predicted trend (blue
- -), predicted confidence interval ($95 \%$) (black --), price's
forecast higher than the predicted trend (green $\bigtriangleup$) ,
price's forecast lower than the predicted trend (blue
$\bigtriangledown$) \label{p3a}}
\end{figure*}
\begin{figure*}[!ht]
\centering {{
\rotatebox{-90}{\resizebox{!}{10.5cm}{%
\includegraphics{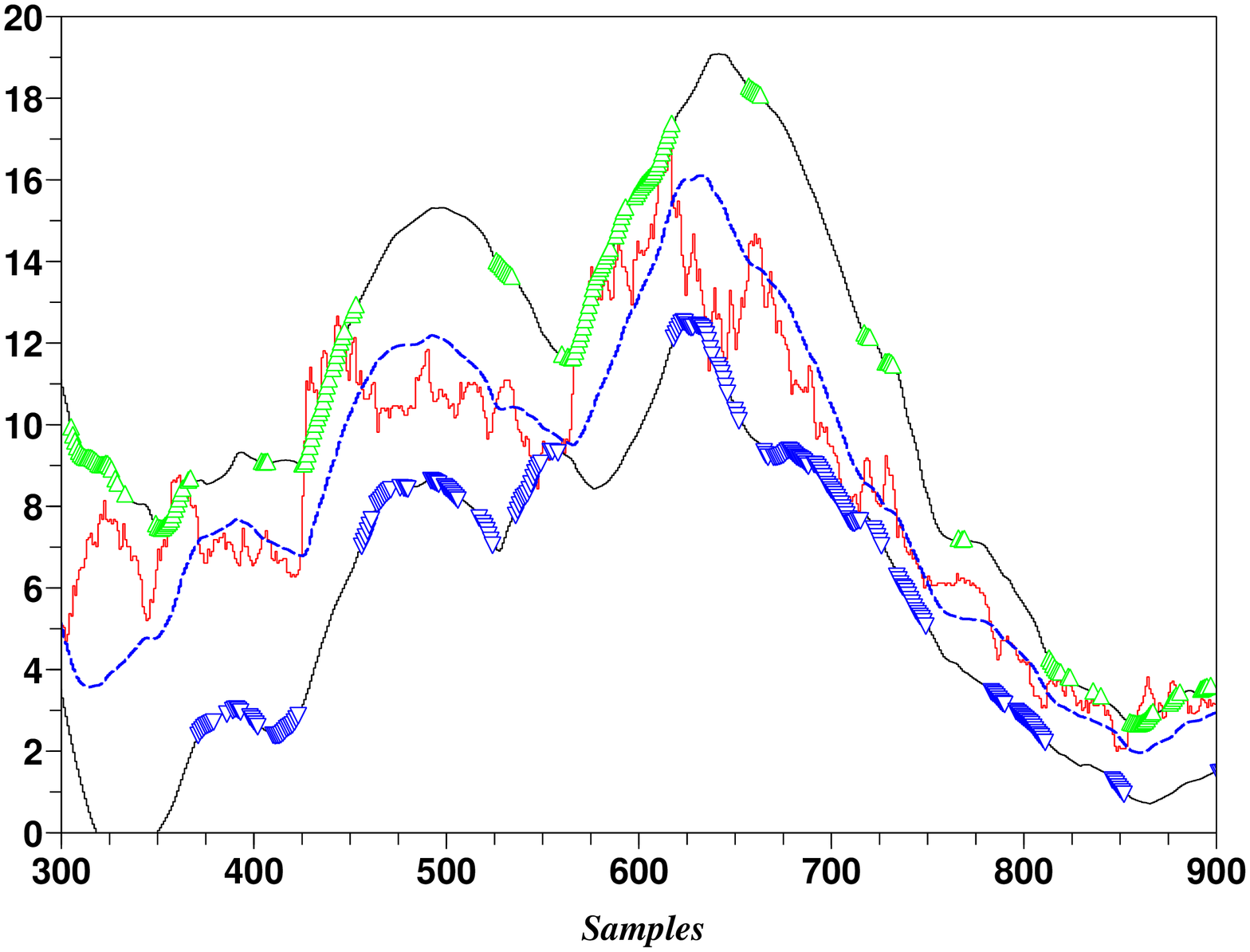}}}}}
\caption{$1$ day forecast -- Zoom of figure \ref{p3a} \label{p3b}}
\end{figure*}

\begin{figure*}[!ht]
\centering {{
\rotatebox{-90}{\resizebox{!}{10.5cm}{%
\includegraphics{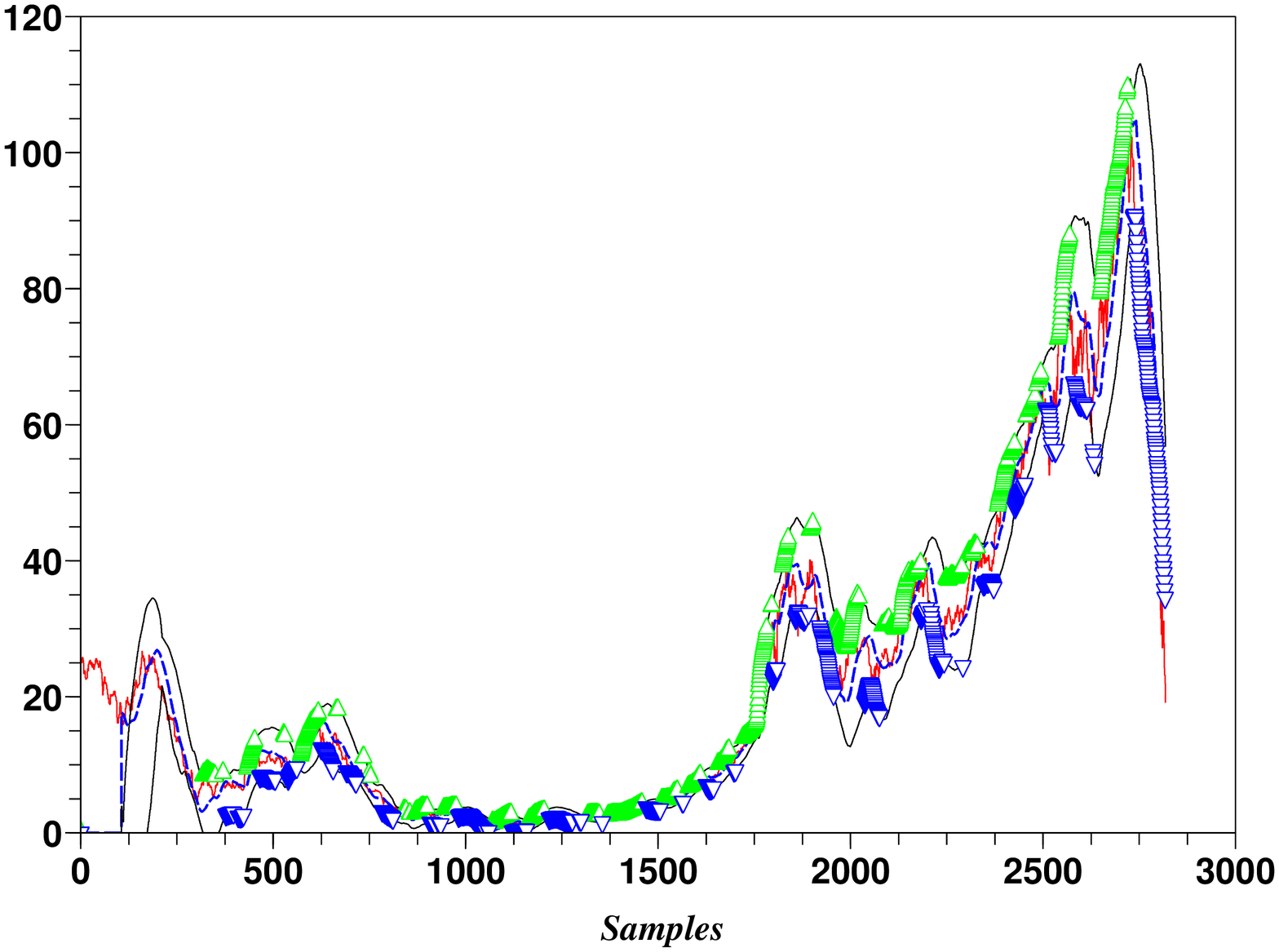}}}}}
\caption{$5$ days forecast -- Prices (red --), predicted trend (blue
- -), predicted confidence interval ($95 \%$) (black --), price's
forecast higher than the predicted trend (green $\bigtriangleup$) ,
value is forecasted as lower than predicted trend (blue
$\bigtriangledown$) \label{p4a}}
\end{figure*}
\begin{figure*}[!ht]
\centering {{
\rotatebox{-90}{\resizebox{!}{10.5cm}{%
\includegraphics{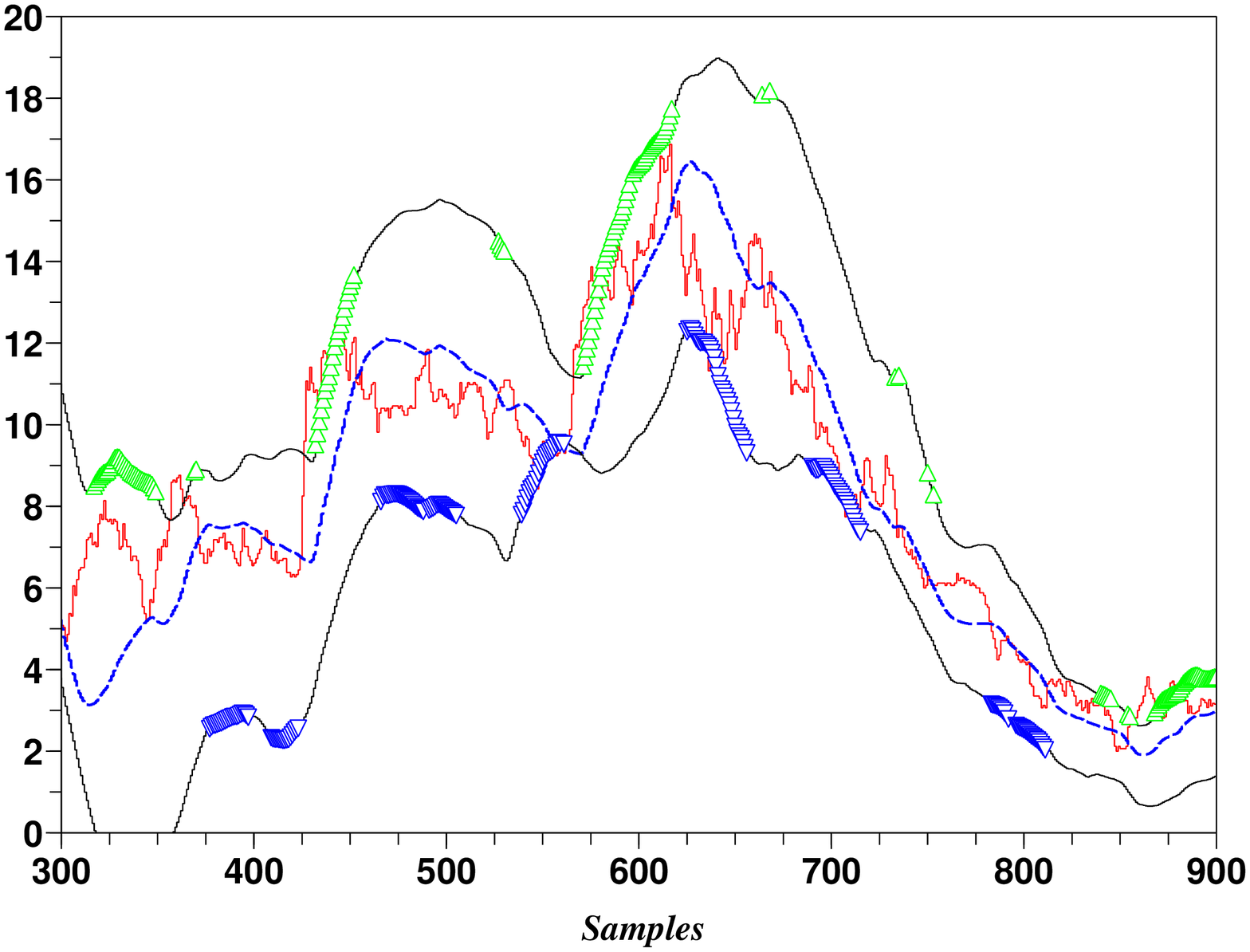}}}}}
\caption{$5$ days forecast -- Zoom of figure \ref{p4a} \label{p4b}}
\end{figure*}

Estimating the first two derivatives yields a forecast of the price
position above or under the trend. The results reported in Figures
\ref{p3a}-\ref{p3b} show for $1$ day (resp. $5$ days) ahead
 $75.69\%$ (resp. $68.55\%$) for an exact prediction, $3.54\%$ (resp. $3.69\%$) without
any decision, $20.77\%$ (resp. $27.76\%$) for a wrong prediction.

\section{Conclusion: probability in quantitative finance}\label{conclusion}
The following question may arise at the end of this preliminary
study on trends in financial time series:
\\ ~ \\
Is it possible to improve the forecasts given here and in
\cite{coventry} by taking advantage of a precise probability law for
the fluctuations around the trend?
\\ ~ \\
Although Mandelbrot \cite{mandel1} has shown in a most convincing
way more than forty years ago that the Gaussian character of the
price variations should be at least questioned, it does not seem
that the numerous investigations which have been carried on since
then for finding other probability laws with jumps and/or with ``fat
tails'' have been able to produce clear-cut results, {\em i.e.},
results which are exploitable in practice (see, {\em e.g.}, the
enlightening discussions in \cite{jondeau,mandel2,sornette} and the
references therein). This shortcoming may be due to an ``ontological
mistake'' on uncertainty:
\\ ~ \\
Let us base our argument on new advances in {\em model-free control}
\cite{sm}. Engineers know that obtaining the differential equations
governing a concrete plant is always a most challenging task: it is
quite difficult to incorporate in those equations
frictions,\footnote{Those frictions have nothing to do with what are
called {\em frictions} in market theory!} heating effects, ageing,
etc, which might have a huge influence on the plant's behavior. The
tools proposed in \cite{sm} for bypassing those
equations\footnote{The effects of the unknown part of the plant are
estimated in the model-free approach and not neglected as in the
traditional setting of {\em robust control} (see, {\em e.g.},
\cite{doyle} and the references therein).} got already in spite of
their youth a few impressive industrial applications. This is an
important gap between engineering's practice and theoretical physics
where the basic principles lead to equations describing ``stylized''
facts. The probability laws stemming from statistical and quantum
physics can only be written down for ``idealized'' situations. Is it
not therefore quite na\"{\i}ve to wish to exhibit well defined
probability laws in quantitative finance, in economics and
management, and in other social and psychological sciences, where
the environmental world is much more involved than in any physical
system? In other words {\bf a mathematical theory of uncertain
sequences of events should not necessarily be confused with
probability theory}.\footnote{It does not imply of course that
statistical tools should be abandoned (remember that we computed
here standard deviations, skewness, kurtosis).} To ask if the
uncertainty of a ``complex'' system is of probabilistic
nature\footnote{We understand by ``probabilistic nature'' a precise
probabilistic description which satisfies some set of axioms like
Kolmogorov's ones.} is an undecidable metaphysical question which
cannot be properly answered via experimental means. It should
therefore be ignored.

\begin{remark}
One should not misunderstand the authors. They fully recognize the
mathematical beauty of probability theory and its numerous and
exciting connections with physics. The authors are only expressing
doubts about any modeling at large in quantitative finance, with or
without probabilities.
\end{remark}

The Cartier-Perrin theorem \cite{cartier} which is decomposing a
time series as a sum of a trend and a quickly fluctuating function
might be
\begin{itemize}
\item a possible alternative to the probabilistic viewpoint,
\item a useful tool for analyzing
\begin{itemize}
\item different time scales,
\item complex behaviors, including abrupt
changes, {\em i.e.}, ``rare'' extreme events like financial crashes
or booms, without having recourse to a model via differential or
difference equations.
\end{itemize}
\end{itemize}
We hope to be able to show in a near future what are the benefits
not only in quantitative finance but also for a new approach to time
series in general (see \cite{coventry} for a first draft).

\end{document}